# Near-Infrared Spectral Monitoring of Pluto's Ices II: Recent Decline of CO and $N_2$ Ice Absorptions

W.M. Grundy[1,2], C.B. Olkin[2,3], L.A. Young[2,3], and B.J. Holler[2,4].

1. Lowell Observatory, 1400 W. Mars Hill Rd., Flagstaff AZ 86001.
2. Remote observer at the Infrared Telescope Facility, which is operated by the University of Hawaii under Cooperative Agreement #NNX-08AE38A with the National Aeronautics and Space Administration, Science Mission Directorate, Planetary Astronomy Program.
3. Southwest Research Institute, 1050 Walnut St. #300, Boulder CO 80302.
4. Laboratory for Atmospheric and Space Physics, University of Colorado at Boulder, 1234 Innovation Dr., Boulder CO 80303.

**ABSTRACT**

IRTF/SpeX observations of Pluto's near-infrared reflectance spectrum during 2013 show vibrational absorption features of CO and $N_2$ ices at 1.58 and 2.15 µm, respectively, that are weaker than had been observed during the preceding decade. To reconcile declining volatile ice absorptions with a lack of decline in Pluto's atmospheric pressure, we suggest these ices could be getting harder to see because of increasing scattering by small $CH_4$ crystals, rather than because they are disappearing from the observed hemisphere.

## 1. Introduction

Near-infrared reflectance spectra of Pluto and Charon obtained on 65 nights at NASA's Infrared Telescope Facility (IRTF) during 2001-2012 were recently presented by Grundy et al. (2013; hereafter Paper 1). The spectra are dominated by vibrational absorptions of methane ice on Pluto's surface (for an overview see Cruikshank et al. 1997 and 2014, and references therein). A few absorptions are caused by ices of nitrogen and carbon monoxide, both overtones of absorptions at longer wavelengths not easily observable from Earth-based telescopes (Owen et al. 1993; Douté et al. 1999). The absorptions of all three ice species are modulated by Pluto's 6.4 day diurnal rotation, enabling constraints to be placed on the longitudinal distributions of the different ices. Over seasonal timescales, varying patterns of insolation are thought to cause sublimation and redistribution of these volatile ices (e.g., Spencer et al. 1997; Trafton et al. 1998). At Pluto surface temperatures of around ~40 K (e.g., Tryka et al. 1994; Lellouch et al. 2000, 2011) all three ices are mobile, but CO and $N_2$ ices are much more volatile than $CH_4$ ice (Fray and Schmitt 2009). CO and $N_2$ ices are also miscible in one another at all concentrations (Vetter et al. 2007), so it is not surprising that their absorptions were found to share similar longitudinal modulations, being stronger by roughly a factor of two when Pluto's anti-Charon hemisphere is oriented toward the observer. The simplest explanation is that the two ices occur together on Pluto's surface. The longitudinal modulations of Pluto's numerous $CH_4$ ice absorptions have lower amplitude and their maximum absorption is associated with the leading hemisphere[1]. This observation suggests that Pluto's $CH_4$ ice has a regional distribution quite distinct from that of more volatile CO and $N_2$ ices, although other scenarios involving regionally varying ice textures or stratigraphic structures could produce comparable longitudinal variability.

Comparing data averaged over 2001-2006 and over 2007-2012, Paper 1 reported that Pluto's CO and $N_2$ absorptions appeared to be declining over time, although the observed change between those two intervals was only marginally significant for the CO absorption band. Two possible interpretations were considered, one involving a static distribution of ices seen from a gradually changing vantage as Pluto moves along its orbit and one involving changing spatial distributions of ices as a result of seasonal volatile transport. The static, geometric interpretation places the source of CO and $N_2$ absorptions at lower latitudes that are contributing progressively less to Pluto's disk-averaged spectrum as the sub-Earth latitude moves away from the equator (equinox occurred during the 1980s and by 2012 the sub-Solar latitude had reached 48°N, with north being defined by the direction of the system angular momentum vector). The volatile transport explanation involves CO and $N_2$ ices sublimating away from the sunlit northern hemisphere. Both factors could be contributing to the observed decline and thus could be difficult to distinguish. One possible way to disentangle the two effects exploits the parallax from Earth's motion around the Sun. Pluto's obliquity is high, so the slow northward progression of the sub-Earth latitude is modulated by the Earth's annual motion, making it possible to schedule a pair of observations during two successive Pluto apparitions such that the sub-Earth latitude and longitude are nearly identical on both dates. Three such geometrically-matched pairs of observations were considered in Paper 1, with little change being observed between the spectrum pairs. This observation was interpreted as favoring the geometric explanation, although the measurement uncertainties were such that a substantial contribution from volatile transport could not be ruled out.

Unlike the CO and $N_2$ absorptions, Pluto's stronger $CH_4$ ice absorptions have been gradually increasing in depth, a trend that has endured for some time (e.g., Grundy and Fink 1996; Grundy and Buie 2001), although it may have slowed in recent years, according to Paper 1. The change has been more dramatic at some longitudes than others. Again, two potential causes were considered. A geometric explanation involved $CH_4$ absorption being especially strong in northern high-latitude regions that are increasingly dominating the disk-integrated spectrum. A volatile transport explanation involved $CH_4$ being progressively concentrated by sublimation loss of more volatile $N_2$ and CO ices. But the geometrically-matched pairs showed declining $CH_4$ band strength, contrary to expectation from either mechanism acting alone. If volatile transport was solely responsible for the long term evolution, we would have expected to see increasing $CH_4$ band strength in the matched pairs, while if Pluto's ices were static we would have expected to see no change in geometrically-matched pairs. Paper 1 proposed one possible way to reconcile the observation with both scenarios: coupling an increase in $CH_4$ absorption due to geometric effects with simultaneous sublimation loss of $CH_4$ from the northern hemisphere or reduction in mean optical path length in $CH_4$ ice due to evolving ice texture.

---

1  Pluto's leading hemisphere is the hemisphere centered on 270°E in right-handed coordinate systems where 0° longitude is defined by the direction toward Charon and the Sun rises in the east. The barycenter of the system is about a Pluto radius above Pluto's surface, so the leading apex is not actually in the center of this hemisphere, but at a longitude of ~300°E.

## 2. Observations

Continuing our long-term monitoring campaign, we observed Pluto and Charon on seven nights during 2013 using the SpeX infrared spectrometer at IRTF (Rayner et al. 1998, 2003). Observing circumstances appear in Table 1. Procedures for data acquisition and reduction were described in Paper 1 and interested readers are referred there for additional details. We used a 0.3 arcsec slit with SpeX's short cross dispersed mode, resulting in wavelength coverage from 0.8 to 2.4 µm with a typical spectral resolution in nightly average spectra of around $\lambda/\Delta\lambda \approx 1600$ where $\lambda$ is wavelength and $\Delta\lambda$ is the apparent full width at half maximum of an unresolved line. Bracketing observations of nearby solar analog star HD 170379 were used to remove instrumental and telluric effects from the spectra, after correcting the star spectra for a slightly hotter than solar effective temperature of ~6500 K. Example spectra are shown in Paper 1, specifically Fig. 1 for individual nightly spectra and Fig. 12 for an average spectrum with absorptions by $CH_4$, CO, and $N_2$ ices labeled.

**Table 1**
Observational Circumstances

| UT date of observation mean-time | Sky conditions | H band seeing (″) | Sub-Earth longitude[a] (°E) | Sub-Earth latitude[a] (°N) | Phase angle (°) | Pluto integration time (min) |
|---|---|---|---|---|---|---|
| 2013/05/21 13$^h$.90 | Clear | 0.4 | 44.9 | 49.7 | 1.17 | 64 |
| 2013/06/01 12$^h$.78 | Partly cloudy | 1.0 | 147.7 | 49.5 | 0.90 | 134 |
| 2013/06/12 13$^h$.56 | Patchy cirrus | 0.8 | 246.0 | 49.3 | 0.59 | 66 |
| 2013/06/26 10$^h$.63 | Clear | 0.7 | 184.1 | 49.0 | 0.20 | 96 |
| 2013/06/27 10$^h$.62 | Clear | 0.6 | 127.7 | 49.0 | 0.17 | 104 |
| 2013/06/30 10$^h$.52 | Clear | 1.0 | 318.9 | 48.9 | 0.11 | 100 |
| 2013/07/27 8$^h$.51 | Clear | 0.5 | 239.9 | 48.4 | 0.76 | 104 |

Table notes:
[a.] Throughout this paper, geometry on Pluto is defined as in previous publications of this author, using a right-hand-rule coordinate system in which north is defined by the direction of Pluto's spin vector and east is the direction of sunrise. Zero longitude is defined by the sub-Charon point. Longitudes and latitudes tabulated here were computed by assuming Pluto's spin state is tidally locked to the Buie et al. (2012) orbit of Charon.

## 3. Evolving $N_2$ and CO Ice Absorptions

As in Paper 1, for each night's Pluto reflectance spectrum we computed equivalent widths for the CO and $N_2$ ice absorption bands by integrating one minus the ratio of the spectrum divided by a continuum function. For the 1.58 µm overtone band of CO ice, we integrated over wavelengths from 1.575 to 1.581 µm, with a linear continuum function being fitted to the data between 1.571 and 1.575 µm and between 1.581 and 1.585 µm. For the 2.15 µm $N_2$ band, we used wavelengths from 2.130 to 2.165 µm for the band, and 2.115 to 2.130 µm and 2.165 to 2.180 µm for the linear continuum model. Pluto's $CH_4$ bands are broader and overlap one another, so we could not compute their equivalent widths. Instead we computed their fractional depths, using the wavelengths in Table 2 of Paper 1. Figure 1 compares the new 2013 measurements of CO, $N_2$, and $CH_4$ ice absorptions with previously published measurements from Paper 1 (Figures 3, 6, and 4 for CO, $N_2$, and $CH_4$, respectively). It is immediately apparent that all seven of the 2013 observations show CO absorptions well below the average behavior seen in the earlier data, indicating a striking decline in absorption by Pluto's CO ice since the earlier observa-tions. The $N_2$ equivalent widths in the middle panel were also lower on average than before, but less obviously so, owing to greater scatter among the measurements for that band. The new 2013 $CH_4$ data show no comparably striking change in the depths of that ice's absorption bands, with the example of the 1.72 µm $CH_4$ ice band being shown in the bottom panel.

We caution that the 2.15 µm $N_2$ band occurs on the shoulder of a much stronger $CH_4$ ice absorption, the variation of which could contaminate the apparent $N_2$ equivalent width. Another source of uncertainty is that the Pluto spectra are contaminated with variable contributions from Pluto's largest satellite, Charon, depending on the nightly location of Charon relative to Pluto as well as the seeing conditions. Charon flux would be most noticeable in the cores of the deepest $CH_4$ absorption bands where Pluto is least bright relative to Charon (see Fig. 1 in Grundy and Buie 2002). We can estimate the maximum possible effect on the observed bands by considering that Charon contributes at most 22% of the projected area of the system (for Pluto and Charon radii of 603.6 and 1151 km, respectively; from Sicardy et al. 2006 and Buie et al. 2010). Considering the relative albedos of Pluto and Charon at the wavelengths of the absorption bands, we can estimate a ±7% maximum contribution to the variability of the CO ice band and ±10% to that of the $N_2$ ice band. In both cases the variability from Charon contamination is much less than the scatter in the data. But for the fractional depth of the 1.72 µm $CH_4$ ice band, we estimate a ±9% maximum contribution from variable Charon contamination, an appreciable fraction of the observed scatter in that band.

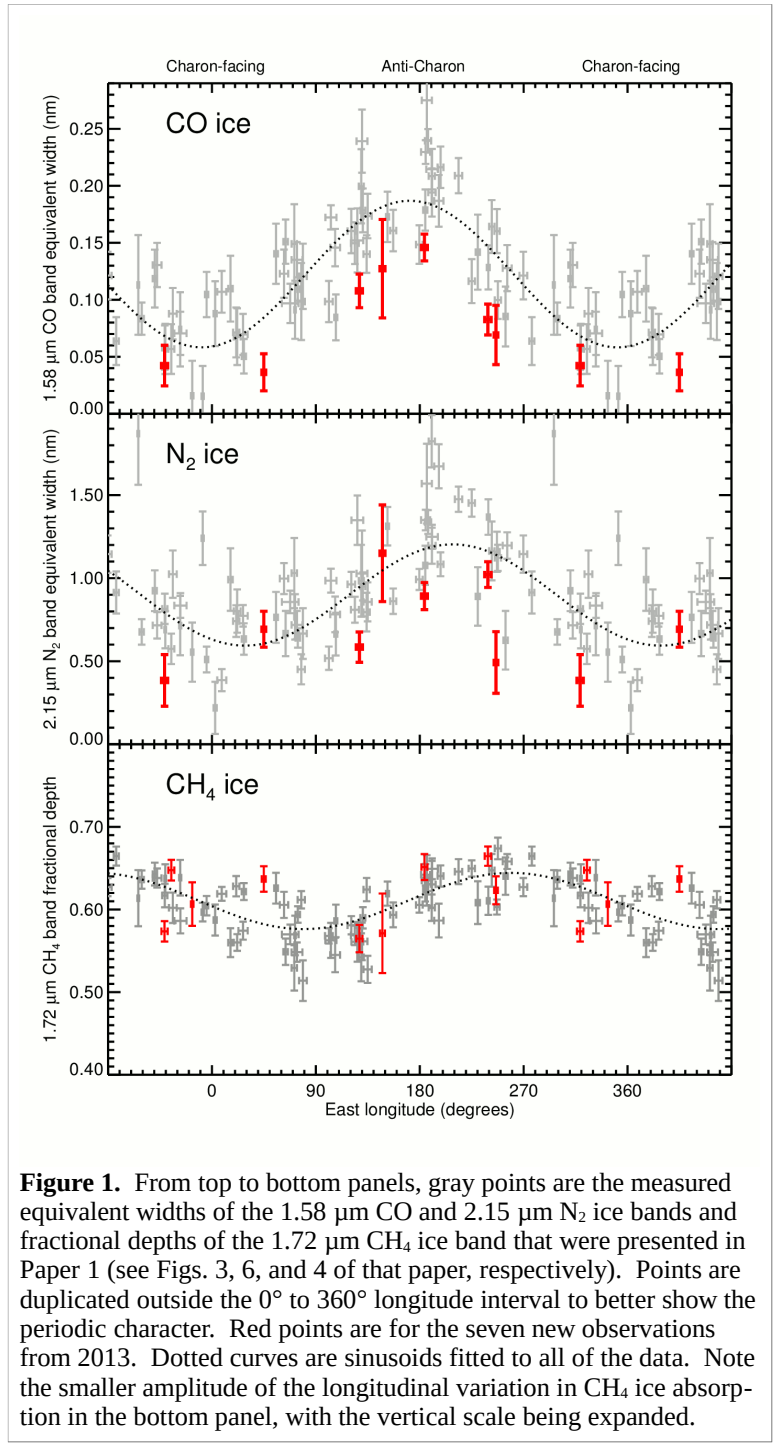

**Figure 1.** From top to bottom panels, gray points are the measured equivalent widths of the 1.58 µm CO and 2.15 µm $N_2$ ice bands and fractional depths of the 1.72 µm $CH_4$ ice band that were presented in Paper 1 (see Figs. 3, 6, and 4 of that paper, respectively). Points are duplicated outside the 0° to 360° longitude interval to better show the periodic character. Red points are for the seven new observations from 2013. Dotted curves are sinusoids fitted to all of the data. Note the smaller amplitude of the longitudinal variation in $CH_4$ ice absorption in the bottom panel, with the vertical scale being expanded.

The reduction in volatile CO and $N_2$ ice absorptions is clearly visible in the 2013 data in Fig. 1, but we wanted to explore the time history of these declining absorptions in greater detail. To do this, we took the sinusoidal fit to the entire data set for each ice absorption band as a func-tion of sub-Earth longitude (the dotted curves in the top two panels of Fig. 1) and refit the constant terms to minimize $\chi^2$ separately to data from each year for which two or more observations were available. This fitted constant term can be interpreted as an average absorption over Pluto's diurnal rotation. Although sinusoids with fixed amplitude and phase are at best a crude approximation of the true functional form of the longitudinal variations of Pluto's CO and $N_2$ ice absorptions, this procedure offers at least some correction for the different longitude sampling from one year to the next. The resulting year-by-year evolution for both ices is shown in Fig. 2. The aver-age CO and $N_2$ equivalent widths are lower in 2013 than in any previous year, although by these measures, 2010 was also a relatively low year. Indeed, it looks as if the decline in $N_2$ ice absorption could have begun as early as 2007 to 2009. By averaging together data from 2007-2012, Paper 1 had blended three years when Pluto still showed relatively strong CO absorption with three having diminished absorption, diluting the signal of declining CO ice absorption. Paper 1 also considered the amplitude of the variation in these ice absorptions, reporting a decrease in amplitude of the CO absorption when data from 2001-2006 were compared with data from 2007-2012. If we fit both the constant and amplitude terms of the sine function to data from each year, the uncertainty in the fitted parameters increases since more parameters are being constrained by the available data, but the year-to-year patterns in the constant term look similar. The amplitudes also appear to be diminishing over time, although there is a lot of scatter. We anticipate doing a more detailed analysis of these changes, including their longitudinal dependence, in a future paper.

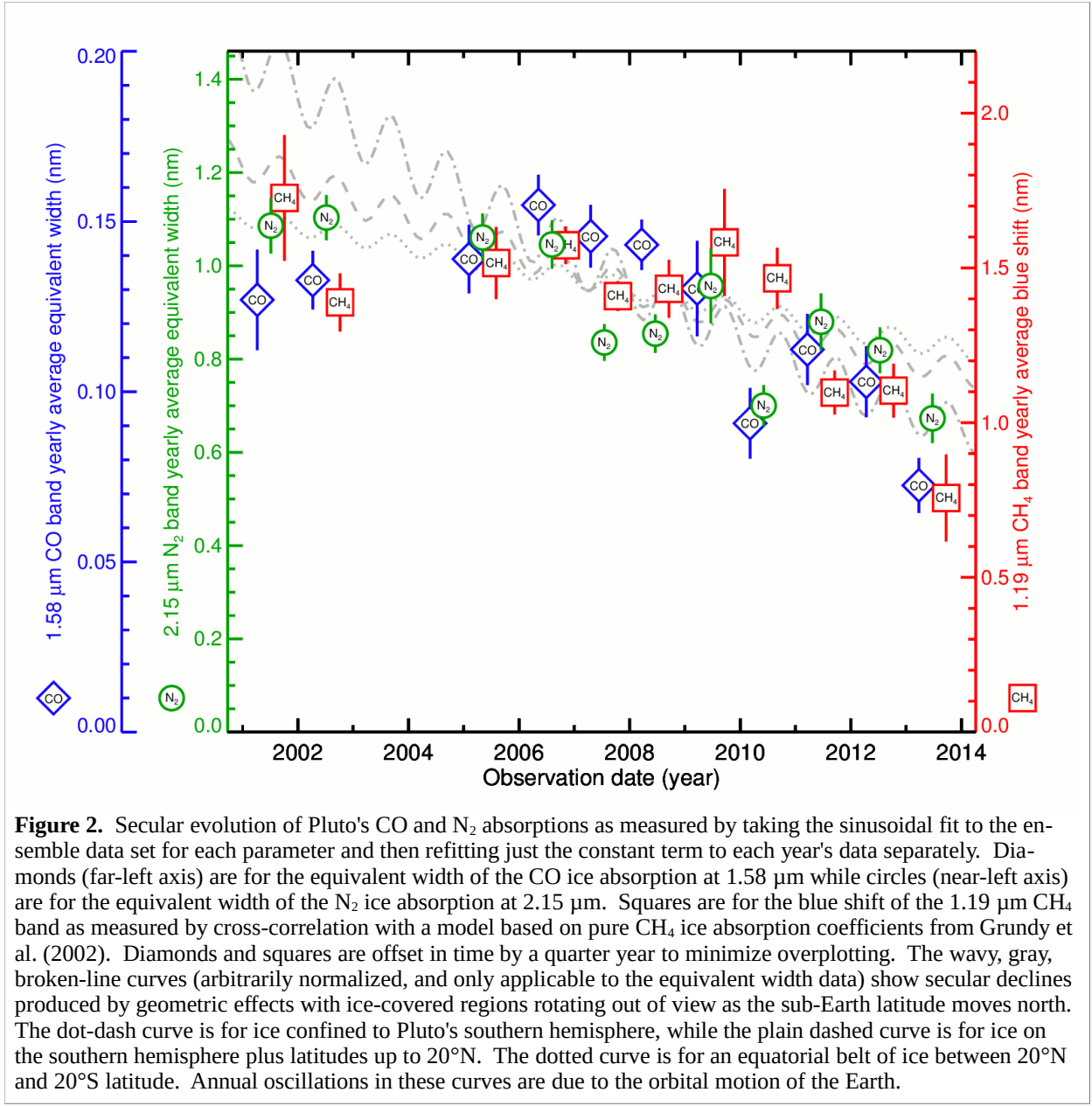

**Figure 2.** Secular evolution of Pluto's CO and $N_2$ absorptions as measured by taking the sinusoidal fit to the ensemble data set for each parameter and then refitting just the constant term to each year's data separately. Diamonds (far-left axis) are for the equivalent width of the CO ice absorption at 1.58 µm while circles (near-left axis) are for the equivalent width of the $N_2$ ice absorption at 2.15 µm. Squares are for the blue shift of the 1.19 µm $CH_4$ band as measured by cross-correlation with a model based on pure $CH_4$ ice absorption coefficients from Grundy et al. (2002). Diamonds and squares are offset in time by a quarter year to minimize overplotting. The wavy, gray, broken-line curves (arbitrarily normalized, and only applicable to the equivalent width data) show secular declines produced by geometric effects with ice-covered regions rotating out of view as the sub-Earth latitude moves north. The dot-dash curve is for ice confined to Pluto's southern hemisphere, while the plain dashed curve is for ice on the southern hemisphere plus latitudes up to 20°N. The dotted curve is for an equatorial belt of ice between 20°N and 20°S latitude. Annual oscillations in these curves are due to the orbital motion of the Earth.

We explored the temporal patterns of spectral changes that could be caused by static ice distributions coupled with the seasonal geometric evolution by generating Pluto models having various simple distributions of an optically active ice and sampling them as they would be seen by an Earth-based observer during the years spanned by our IRTF observations. The temporal evolution of a few examples are shown in broken, gray curves in Fig. 2. In one of these models, the ice was confined to Pluto's southern hemisphere. Another was similar, but with the ice extending up to 20°N latitude. In a third, the ice occurred in an equatorial belt from 20°S to 20°N latitude. Although these sorts of very simplistic models can match the recent rate of decline, or the average decline over the past 13 years, none of them was able to replicate a relatively steady-state period during 2001-2006 followed by an accelerating decline of CO and $N_2$ ice absorptions in subsequent years. These tests do not prove that a more complex distribution of static ices could not do that, but it is not obvious that strictly geometric factors could match an accelerating pace of spectral changes.

Pluto's $CH_4$ bands can provide a separate indicator of $N_2$ ice abundance. Close examination of $CH_4$ bands shows they consist of two components, interpreted as being due to absorption by $CH_4$ ice itself plus a blue-shifted version of the same band attributed to $CH_4$ diluted in $N_2$ ice (e.g., Quirico and Schmitt 1997; Douté et al. 1999). Tegler et al. (2010, 2012) pointed out that two such components can co-exist under conditions of thermodynamic equilibrium if $CH_4$ exceeds its solubility limit in $N_2$ (about 5% molar concentration of $CH_4$ in $N_2$ ice at 40K) and that the two phases present under such circumstances would be $N_2$ ice saturated with $CH_4$ and also $CH_4$ ice saturated with $N_2$ at its solubility limit (about 3% $N_2$ in $CH_4$ ice at 40K). The solubility limit compositions are functions of temperature (e.g., Prokhvatilov and Yantsevich 1983; Lunine and Stevenson 1985), but assuming thermodynamic equilibrium at a given temperature, the ratio of $N_2$ to $CH_4$ dictates the relative abundances of the two phases. A larger shift indicates more of the $CH_4$ is in the shifted, $CH_4$ diluted in $N_2$ component, and thus there must be a higher $N_2$ abundance. We used cross-correlation to measure the apparent shifts of the $CH_4$ bands in our Pluto spectra and found that they were smaller in 2013 than in previous years. The square points in Fig. 2 show how the yearly evolution of the shift of an example band is somewhat similar to the yearly evolution of the $N_2$ and CO ice equivalent widths, boosting our confidence in the recent decline in absorption by these volatile ices.

The rates of seasonal decline in the three measures of Pluto's volatile ices in Fig. 2 all give the appearance of accelerating in recent years. Our existing data are not yet sufficient to distinguish definitively between accelerating and constant rates of change, but they make spectroscopic monitoring of the Pluto system over the next few years look especially timely. We are not convinced that apparent fluctuations in Fig. 2 from one year to the next are necessarily real, since as discussed earlier, our correction for Pluto's diurnal variation is imperfect, and the temporal sampling was essentially random each year, and relatively sparse (see Table 1 of Paper 1). For instance, we obtained no observations during 2010 at sub-Earth longitudes near 0°. Pluto could have appreciable absorption variations from year to year, but our observations have been too infrequent to make a convincing case for them.

## 4. Discussion

Recent decreases in Pluto's $N_2$ and CO equivalent widths are attributable to some combination of volatile transport removal and/or a static distribution coupled with seasonally changing geometry. But both mechanisms seem inconsistent with data from stellar occultations during the same time period, showing that Pluto's atmospheric pressure at a reference radius of 1275 km has not declined (Bosh et al. 2013; Olkin et al. 2014). On Pluto, the atmospheric pressure at the surface is supported by vapor-pressure equilibrium over the $N_2$ ice, and the $N_2$ ice temperature is equalized over the surface through the atmospheric transport of latent heat (e.g., Spencer et al. 1997; Trafton et al. 1998). Thus, the pressure depends on the absorbed insolation averaged over the $N_2$ ice. If the observed decreases in $N_2$ and CO equivalent widths and $CH_4$ shifts were due to a decrease in the projected area of the volatile ices as seen from Earth, regardless of whether the decrease was due to changing geometry or to volatile transport, there would be a corresponding decrease in solar illumination of the ice and thus in its net absorption of radiation. Less absorption would in turn lead to a decrease in surface pressure, contrary to the occultation results. A concurrent reduction in the albedo or emissivity of the $N_2$-rich ice could increase net energy absorption, counteracting decreasing sunlit projected area. However, such changes ought to have other observable consequences that would need to be reconciled with Pluto's evolving visible reflectance lightcurve and colors (e.g., Buratti et al. 2003; Buie et al. 2010) as well as thermal emission (e.g., Lellouch et al. 2000, 2011).

The phase behavior of binary mixtures of $N_2$ and $CH_4$ ices mentioned earlier offers a possible understanding of how the observed declines in absorption by Pluto's volatile CO and $N_2$ ices and in the shifts of the $CH_4$ bands could be reconciled with the absence of observed reductions in atmospheric pressure, $CH_4$ ice absorption, and photometric brightness. Insolation-driven sublimation of $N_2$ from the $N_2$-rich phase in the $N_2$:$CH_4$ binary system would force a proportional amount of $CH_4$ out of the $N_2$-rich phase in order to maintain its composition as dictated by the solubility limit. The exsolved $CH_4$ would contribute to growth of existing $CH_4$-rich ice crystals and/or nucleate new ones. The balance between nucleation of new crystals and growth of existing ones would depend on several unknown factors, including their spacing and the diffusion rates of $CH_4$ and $N_2$ ices through one another. In the limit of no diffusion, $CH_4$ would accumulate exactly where sublimation removed $N_2$, leading to the Trafton et al. (1997) "detailed balancing" scenario where $N_2$ sublimation is choked off by the build up of a $CH_4$ rich layer. But unless diffusion rates are extremely slow, thermodynamic equilibrium suggests $CH_4$ crystal growth could occur not just at the surfaces of sublimating grains, but throughout the optically active uppermost several cm of ice where sunlight is absorbed (e.g., Grundy and Stansberry 2000). The $CH_4$ crystals seem likely to be small and close-spaced compared with the centimeter or larger spatial scales of scattering within $N_2$ ice estimated from multiple scattering radiative transfer models of Pluto's $N_2$ ice absorption (e.g., Douté et al. 1999; this size is often described as a "grain size", although it could equally be envisioned as a characteristic spacing between scatterers such as voids or fractures in a more compacted material). By producing additional scattering, the growth and/or nucleation of small $CH_4$-rich crystals in response to sublimation of $N_2$ could reduce the mean free path of photons in the $N_2$ ice, causing the observed $N_2$ absorptions to decrease without reducing the geographic extent of sunlit $N_2$ needed to support the atmosphere. The $CH_4$ bands would appear less shifted because the fraction of $CH_4$ in the $CH_4$-rich phase had increased. For the depths of the $CH_4$ ice absorptions not to change much (see the bottom panel of Fig. 1) requires the quantity of $CH_4$ encountered by photons that scatter out of the surface to remain similar despite the increase in scattering and decrease in mean free path. We suspect that the depths of the $CH_4$ bands can be maintained if the additional scattering is caused by $CH_4$-rich crystals, but detailed models beyond the scope of this paper are needed to determine if this speculation works quantitatively. That effort is already in progress, using new laboratory optical constants of the two ice phases, each with the other species at its solubility limit (Protopapa et al. 2013). It is also worth noting that this discussion should be based on phase stability in the $N_2$:$CH_4$:CO ternary system. But the necessary laboratory studies of that system are not available, so for now we are restricted to considering only the $N_2$:$CH_4$ binary system.

NASA's New Horizons probe will fly through the Pluto system in 2015, sending back a wealth of data from its sophisticated suite of instruments (e.g., Young et al. 2008). However, the spectra presented in this paper show Pluto's surface could be evolving on time scales that are short compared with Pluto's 2.5 century heliocentric orbit, but long compared with the duration of the New Horizons flyby. It is crucial to continue regular monitoring of Pluto from Earth-based observatories to help fit the detailed snapshot from New Horizons into the longer term seasonal evolution of Pluto's volatile ices and atmosphere. A combination of observing techniques including near-infrared spectroscopy, visible photometry, frequent stellar occultations, and thermal infrared radiometry can provide powerful constraints on Pluto's seasonal evolution.

## Acknowledgments

The authors gratefully thank the staff of IRTF for their assistance with this project, especially W. Golisch, D. Griep, E. Volquardsen, and B. Cabreira. The work was funded in part by NASA Planetary Astronomy grant NNX09AB43G. We wish to recognize and acknowledge the significant cultural role and reverence of the summit of Mauna Kea within the indigenous Hawaiian community and to express our appreciation for the opportunity to observe from this special mountain. We thank the free and open source software communities for empowering us with key software tools used to complete this project, notably Linux, the GNU tools, MariaDB, LibreOffice, Evolution, Python, and FVWM. Finally, we thank two anonymous reviewers for their constructive suggestions that helped to improve this paper.